\documentclass[12pt]{iopart} 
\usepackage{hyperref}
\usepackage{cite}
\usepackage{color}
\usepackage{tabularx}
\hypersetup{breaklinks=true}
\begin{document}

\title{Challenges and opportunities for AI to help deliver fusion energy}

\author{Adriano Agnello}
\address{STFC Hartree Centre, SciTech Daresbury, Warrington, WA4 4AD, UK}
\author{Helen Brooks}
\address{UK Atomic Energy Authority, Culham Campus, Abingdon, Oxfordshire, OX14 3DB, UK}
\author{Cyd Cowley}
\address{digiLab Solutions, Generator Hub The Gallery, Kings Wharf, Exeter, EX2 4AN, UK}
\author{Iulia Georgescu}
\address{Institute of Physics, 37 Caledonian Road, London, N1 9BU, UK}
\author{Alex Higginbottom}
\address{Zenithon AI,  1111B S Governors Ave, Dover DE 19904, USA}
\author{Richard Pearson}
\address{Eindhoven University of Technology, 5600 MB Eindhoven, Netherlands}
\author{Tara Shears}
\address{Oliver Lodge Laboratory, University of Liverpool, Liverpool,  L69 7ZE, UK}
\author{Melanie Windridge}
\address{Fusion Energy Insights, Culham Innovation Centre, Culham Campus, Abingdon, Oxfordshire, OX14 3DB, UK}

\vspace{10pt}

\begin{abstract}
There is great potential for the application of AI tools in fusion research, and substantial worldwide benefit if fusion power is realised. However, using AI comes with its own challenges, many of which can be mitigated if responsible and robust methodologies are built into existing approaches. To do that requires close, long-term collaborations between fusion domain experts and AI developers and awareness of the fact that not all problems in fusion research are best tackled with AI tools. In April 2025, experts from academia, industry, UKAEA and STFC discussed how AI can be used to advance R\&D in fusion energy at the first edition of The Economist\textit{FusionFest} event. This Perspective is an expanded and updated summary of the round table discussion, providing more context and examples.  

\end{abstract}

\section{Introduction}

Nuclear fusion promises abundant and sustainable power generation, but to achieve this, fusion reactors are needed. These complex machines operate under extreme conditions and, therefore, their construction and operation require advances in materials development, engineering, real-time control mechanisms, and remote handling. These challenges are considerable and often computationally-intensive to address (see, for example, ref.~\cite{garcia2025challenges}).

Artificial intelligence (AI) offers a new approach to tackling some of the research problems in fusion. AI methods work by identifying patterns and correlations between variables in datasets. Applications of AI may involve pattern recognition, image interpretation, or real-time decision-making. AI tools could offer transformational improvements in data analysis, potentially surpassing the performance of traditional approaches, and even enabling investigations that were previously considered impossible.

Advances in AI over the past five years have made these methods more accessible, powerful and ubiquitous than ever before (in particular deep-learning and foundation models). It is natural to ask what can AI do to advance fusion, and what challenges need to be overcome to realize its full benefits. These two questions formed the focus of a roundtable discussion hosted by the Institute of Physics at the \textit{FusionFest} Economist event in April 2025. The event brought together experts from academia, industry, UK Atomic Energy Authority (UKAEA) and Science and Technology Facilities Council (STFC) to discuss the current context, opportunities, and challenges, whose perspectives are explored here. Broad consensus of the involved specialists formed around three core challenges: the necessity to match adequate AI methods to fusion problems; the centrality of data limitations; the importance of trustworthiness throughout model development and deployment.

This Perspective is not intended as an exhaustive review of AI for fusion, but rather to highlight the potential to use AI in fusion research and outline longer-term ambitions, arguing that many of the challenges AI presents can be mitigated if responsible and robust methodologies are built into the existing current approaches. This Perspective article aims to stimulate further discussion and collaborations between academia and industry, fusion experts and AI developers.

This Perspective article is structured as follows: first, we introduce the context from which this article originates, then we highlight trends in the use of AI methods in fusion research and recent developments in the field throughout 2025. Next, the challenges and opportunities of using AI in fusion research are discussed. We look at industry-academia collaborations and, before concluding, dive into more detail with a case study on the role of AI in solving the materials challenge in fusion.

\section{Background}

The \textit{Fusion Fest 2025} roundtable discussion was convened in the context of two recent reports: the Institute of Physics (IOP) report\cite{IOP} `Physics and AI: A physics community perspective’ and the Clean Air Task Force (CATF) report\cite{CATF} `A survey of artificial intelligence and high-performance computing applications to fusion commercialization'. These publications outlined the complex relationship between physics, fusion energy, and AI.

Since the publication of these reports, several review articles have appeared. For example, ref.~ \cite{garcia2025challenges} discussed the challenges and opportunities in exascale fusion simulations and ways in which AI methods can contribute to solving the former. Section 6 in ref.~\cite{summers2026roadmap} overviewed fast machine learning for fusion simulation, optimisation, and control. A collection of papers on machine learning applications for fusion was published in the \textit{Journal of Fusion Energy} \cite{rea2025}. Also a review on machine learning and Bayesian inference in nuclear fusion research \cite{pavone2023} appeared in this journal in 2023.

Like AI for science more broadly \cite{wang2023scientific}, AI for fusion research is a fast-growing area, which is reflected in the number of preprint articles cited here. Throughout this Perspective, we provide a selection of examples that, while not meant as an exhaustive coverage of the published literature, are illustrative of the consensus that was formed at the roundtable, as well as of ongoing developments and ambitions.

\section{Established and Emerging AI Approaches in Fusion Research}
\label{sec:methods}
Thus far, the development of machine learning (ML) techniques for fusion has mainly involved more mature, traditional approaches. Methods such as random forests, Gaussian processes (GPs) \cite{williams1998prediction}, and convolutional neural networks (CNNs) have seen increasing application in areas such as: disruption prediction \cite{DiiiD2025}, turbulent transport modelling \cite{Battye}, and plasma control \cite{zanisi2024efficient,Pfau2025,TCV2024}. However, in the last five years, there has been growing momentum in developing foundation models for physics, with limited exploration for fusion specifically. To define what is meant by a `foundation model for physics' we adopt the approach proposed by Choi et al. \cite{choi2025defining} with a focus on generality, reusability, and scalability, that is: models that are pre-trained on broad data and can be adapted to a wide range of downstream tasks through fine-tuning or in-context learning.

The standard motivation for surrogate modelling in physics and engineering is to accelerate the time to results—where a traditional numerical solver may take hours on a High-performance computing (HPC) cluster. ML surrogate models can achieve inference in milliseconds \cite{gopakumar2024plasma,Agnello2024,noh2025fpl}. This acceleration could enable rapid parameter sweeps across the design space, address the systems integration challenge of computational intractability when interfacing simulations for different components, and provide greater accuracy than the frequently used scaling laws or reduced-order models. However, the challenge of this approach lies in the generality and reusability. With traditional methods (including CNNs, standard multi-layer perceptrons, or GPs), although the surrogate model provides accurate results within the training distribution, it may offer limited utility outside this domain. In some application cases, e.g., for targeted simulations or experiments, by the time a dataset for training such a surrogate is developed, the specific answer being sought may have already been found through conventional means. This limitation motivates the development of models that can generalise across parameter space through several approaches: (i) training on large, broad datasets; (ii) algorithmic developments that enable zero-shot or few-shot fine-tuning; (iii) hybrid models combining numerical approaches with ML; and (iv) active learning pipelines, such as Bayesian optimisation, for improved simulation efficiency. These approaches are not mutually exclusive and optimal results may require combining all of the above. However, approaches (i) and (ii) have not been fully pursued in fusion to date.

Below we provide some examples of ongoing development and longer-term promising direction, cautioning that this is not an exhaustive list due to the fast evolving field of AI and of its applications.

\textbf{Physics-Informed Architectures.}
Physics-informed neural networks (PINNs) and physics-informed neural operators offer a promising direction for incorporating domain knowledge into foundation models. Recent work on the TurbulAI-Fusion framework \cite{adepu2025ai} demonstrates how physics-constrained loss functions can improve both accuracy and generalisation when modelling plasma turbulence. By embedding conservation laws and governing equations directly into the learning process, these approaches can produce models that remain physically consistent even when extrapolating beyond training distributions.

\textbf{Neural Operators.} These\cite{kovachki2023neural} represent a class of architectures for fusion foundation models. Unlike traditional neural networks that learn mappings between finite-dimensional vector spaces, neural operators learn mappings between infinite-dimensional function spaces, making them inherently discretisation-invariant. This property is crucial for fusion applications where simulations operate at varying resolutions and where transfer between simulation codes and experimental data is desirable.
The Fourier Neural Operator (FNO) \cite{li2020fourier} is an early leading architecture in this space, demonstrating six orders of magnitude speedup over traditional magneto-hydrodynamic (MHD) solvers while maintaining high accuracy on plasma evolution prediction tasks \cite{gopakumar2024plasma}. Multi-variable extensions handle coupled partial differential equations (PDEs) governing plasma dynamics, capturing the mutual dependence between field variables such as density, electric potential, and temperature \cite{gopakumar2024plasma,carey2025data}. These models have been successfully applied to predict plasma evolution using experimental data from the MAST tokamak. More recently, sparsified time-dependent FNOs (ST-FNO) have demonstrated over 100× memory reduction when applied to extended-MHD and gyrokinetic codes \cite{rahman2024sparsified}.

\textbf{Multi-Modality Foundation Models.} Multi-modal foundation models can integrate heterogeneous data sources common in fusion research: diagnostic time series, imaging data, simulation outputs, and text documentation. Recent work \cite{churchill2025ai} highlights the potential for models to be pre-trained on unlabelled diagnostic data and fine-tuned for tasks such as identifying plasma instabilities and oscillation modes. This approach could enable automated metadata generation and rapid event classification.
\textbf{Hybrid and Multi-Fidelity Architectures.}
Foundation models need not replace high-fidelity simulation entirely, but can serve as components within hybrid frameworks. Neural networks paired with physics-based plasma simulations can achieve accurate predictions during challenging operational phases such as ramp-downs, using only a few hundred experimental pulses for training \cite{wang2025learning} . This hybrid approach addresses data scarcity (see also the discussion in Section 5) by allowing the physics model to provide structural constraints while the neural network learns residual corrections. Multi-fidelity approaches, where models are first trained on abundant low-fidelity data before fine-tuning on limited high-fidelity samples, have also shown promise \cite{Holt2024}.

\section{Recent Developments Throughout 2025 and 2026}
%{What has changed since the publication of the CATF report in November 2024 regarding AI capabilities or fusion activity and requirements?}

% The pace of change has accelerated dramatically. On the AI front, the emergence of models such as DeepSeek and other new LLMs highlights the need for an ecosystem that can rapidly adapt to evolving models. However, this raises questions about trust: how do we trust these models we are using and the data they have been trained on? How can we make decisions that we can trust? In AI the general sentiment shifted from optimism, to we want to see actual results, measurable advances.
% Geopolitical shifts are also influencing the regulatory environment. The outcome of the US elections, for example, will impact AI safety, IP protection, and patents—raising challenges for developing industrial solutions under uncertain conditions. 

Since the publication of the CATF report in November 2024, the field has been evolving in three main directions: the development of techniques \cite{Shukla2025}, the expectations on technology readiness level (TRL) \cite{IAEA2025}, and the policy and regulatory landscape. Note that due to the fast-pace at which the field is evolving, many cited references are preprints.

Multi-Modal Foundation Models (see Section 3) are being developed to deal with heterogeneous datasets, multi-modal training data (e.g., from 2D and 1D experimental data and multi-dimensional simulation data) and technical knowledge elicitation (incorporating information from domain experts, e.g. physics priors). Besides language models, architectures for in-context learning have been applied to tabular and time-series data \cite{Auer2025,Chronos2025,TabPFN2025,TabICL2025,TSPulse2025,Mitra2025}. To deal with the specific challenges of multi-frequency and multi-fidelity data from tokamak discharges, the TokaMind foundation model has been recently published \cite{TokaMind2026}, together with the TokaMark \cite{TokaMark2026} suite of benchmarks and tools to pre-process FAIR-MAST data \cite{FAIRMAST2024} and a clear breakdown of relevant tasks that fusion foundation models fine-tuning should be suited to. The performance of language models has benefited from different training strategies (e.g., \cite{R1v2024,R1Yuan2025}), and Retrieval-Augmented Generation (RAG) \cite{RAG2020} is being applied for seamless interaction with the available scientific literature (e.g., \cite{Loreti2025} provides example architectures for magnetic-confinement fusion).

% \textcolor{blue}{[on FMs, we should insert references to TTM, TSpulse, Chronos-Bolt, TabbPFN, TabICL, Mitra] [we should also spend a sentence on later LLM releases, DeepSeek] [we may cite Loreti et al. as an example of RAG for fusion]}

Expectations on TRL are shifting towards deployment of AI models that can concretely demonstrate real-time applicability. For example, turbulent transport surrogates are available as components of software suites for integrated-modelling of plasmas \cite{zanisi2024efficient}. AI tools are also used to aid complex simulations, to guide the convergence to a solution, or replace expensive subroutines, through multi-fidelity and reduced-order models (e.g., \cite{FPlNet2025,Howard2025}).%\textcolor{blue}{[FPL-net, PINNs, NOps, 10.1088/1741-4326/ad8804]}. 
On the side of control, which requires sub-millisecond latency, AI-based solutions are used more widely in the control room. For example, after the first demonstration in 2022 \cite{degrave2022magnetic}, reinforcement learning has been since demonstrated on multiple tokamaks %\textcolor{blue}{[cite TCV, DiiiD, KSTAR]}
\cite{Pfau2025,TCV2024,DiiiD2025,Wu2025} also for design of experiments and safe rampdown \cite{Bustos2025,wang2025learning}, alongside lightweight models to identify disruption precursors \cite{Yang2025} and divertor detachment \cite{Yu2025,Chen2026}. With these developments, there is now increased focus on the routine deployment of AI in safety-critical operations, and on possible tradeoffs between explainability vs accuracy, as well as in general ensuring sufficient generalisation beyond the operational regimes that have been already explored in experiments or that e.g. the RL agents have been specifically trained on. As real-time control is now incorporating AI solutions, there is an increased need for simulation code that can be seamlessly integrated with those, both for training and for control validation (e.g. \cite{PentlandFPDT, PentlandDynVal}).
% \textcolor{blue}{[At possibly the highest TRL, add a sentence here on AI being used in experiment design (\cite{Bustos2025},\cite{wang2025learning}) and real-time control (\cite{Wu2025} \cite{Chen2026}) ]}

Real-time control already requires validation today, which is done via simulation. This will remain also with control strategies that are discovered via AI. Systematic uncertainties can compound if the simulations themselves are accelerated by replacing some components with surrogates, but already now the simulations for control validation are only an approximate description of the full relevant physics, and in fact surrogates of more complete physical processes may make the control-validation simulations more accurate in describing the processes that should be monitored in real time.

While present-day tokamaks are mainly experiment-oriented machines, to explore new regimes and solutions (e.g., in confinement, heat-facing components, tritium breeding), substantial activity to develop AI methods to benefit larger machines like ITER (an international nuclear fusion research and engineering project designed to demonstrate the feasibility of fusion power) is happening, with targeted funding from EUROfusion (a consortium of national fusion research institutes located in the European Union, the UK, Switzerland and Ukraine). In this context, there is increased focus on the interpretability of the models, and on the robustness and reliability of large language model (LLM) assistants. In line with the spirit of the FAIR-MAST releases \cite{FAIRMAST2024}, the IAEA datalake effort \cite{SinghDahle2026} aims to provide access to multi-machine data for wider development of AI models by the global community, to aid model generalisation and streamlined development through common standards, extending across magnetic-confinement and inertial-confinement frameworks. The field of fusion is also placing greater emphasis on smaller, lightweight models where feasible \cite{zhou2023mini}.

On the policy and regulatory side, approaches in major Western countries partially align on different aspects. After the AI safety summit in Bletchley Park (in November 2023), the United Kingdom has continued international coordination with the United States and European Union and established its own initiatives throughout 2024 and 2025. The AI Security Institute\footnote{ \url{https://www.aisi.gov.uk/about} } is mostly dedicated to safety and security aspects of AI. The AI Opportunities Action Plan \cite{ai_growth_zone} and the UK Industrial Strategy \cite{uk_industrial_strategy} both refer to fusion as one sector of application of frontier AI, both in scientific research and in clean energy development. Fusion energy is prominent in the UK \textit{AI for Science Strategy} \footnote {\url{https://www.gov.uk/government/publications/ai-for-science-strategy}} and in the US \textit{Genesis} mission\footnote{ \url{ https://genesis.energy.gov/}}. The EU funded the establishment of AI Factories \footnote{\url{https://digital-strategy.ec.europa.eu/en/policies/ai-factories}}, including the development of common European Data Services\footnote{https://digital-strategy.ec.europa.eu/en/policies/data-spaces} and workflows for AI development and deployment on the most powerful supercomputers\footnote{https://ai4eosc.eu} \cite{plociennik_2024_11485073}, while also identifying criteria to subject frontier models to increased scrutiny in the AI Act  (finalized in early 2025, \cite{EU_AI_act,AI_act_commentary}), and encouraging adhesion of major laboratories and companies to the AI Pact for responsible innovation. 

In the US, while the impact of AI for innovation has been transversally recognised, the approach to regulating its development and deployment has been less stringent. The EO-14110 \footnote{\url{https://www.federalregister.gov/d/2023-24283}} instructed Agencies to implement provisions on security, privacy and intellectual property, while it did not directly aim to regulate the development and wider adoption of AI solutions that may pose high risks, but rather their procurement and monitoring (for labour rights, privacy and IP) by Federal agencies. EO-14110 was rescinded after November 2024, with a focus change in the US AI Action Plan \cite{US_AI_Actionplan} to targeted investment and less regulation, except for security (in defense and healthcare) and LLM procurement within Federal agencies, together with action to fund US development of exportable hardware and software for AI, and provisions to limit regulatory action by individual States. Some continuity has remained on the side of expediting permits and leasing of Federal land for infrastructure such as Data Centres, across the EO-14141 \cite{EO14141} and the EO-14318 \cite{EO14318}.

As AI becomes more prevalent and foundation models become more complex, the use of computational resources and energy by AI becomes an ever more significant factor in AI growth. In fact, by 2030 it is expected AI may consume 945 terawatt-hours (TWh) of power, similar to the current electricity consumption of Japan \cite{IEA2025EnergyAI}. To prepare for this demand, the UK government has announced its intentions to create AI growth zones \cite{ai_growth_zone}, the first of which will be located on the Culham Campus in Oxfordshire, also home to the UKAEA. This represents a decisive step that will likely attract significant investment. The recently signed Memorandum of Understanding on the \textit{Technology Partnership Deal} \cite{TPD_MOU} (17/09/2025) establishes civil nuclear fusion and AI, alongside quantum computing, as the critical technologies for strategic UK--US collaboration.

\section{Challenges}
%% Challenges / Limitations

Before discussing the opportunities of AI in fusion, it is important to highlight some of the most pressing challenges. 

\subsection{Data Challenges}
An AI model with predictive capabilities needs to be trained on data at scale. Although the pre-training of large models is done on general data, the fine-tuning yields the real performance in downstream tasks. The challenge is how to deal with not having all the necessary data for pre-training or fine-tuning, which often holds true in fusion research. The data challenges for AI in fusion can be split into three broad categories, including data existence, data quality, and data availability.

In terms of the first category, there are numerous reasons why certain datasets could be lacking, or entirely non-existent.  The first is \textit{sparse experimental coverage of the parameter domain}. Fusion experiments are costly and time-consuming to commission, thus measurements may be limited to a specific subset of the relevant operational conditions. For example, there are currently few experimental nuclear facilities capable of producing neutron fluxes having both the intensity and energy spectra representative of fusion power plants \cite{knaster2014ifmif}. Such a lack of data makes it extremely challenging to predict the degradation of materials under irradiation in the fusion environment (the use of AI in helping advance fusion materials is discussed as a case study in Section 8). Related to this is the problem of \textit{variety}. For example, data from plasma experiment campaigns may not adequately cover the range of operational parameters, but rather consist of small variations around a relatively small number of scenarios.

Another reason is \textit{fidelity and ground-truth}. In the context of surrogate model development, simulation data is used as a proxy for experimental data. However, even a single direct numerical simulation may require substantial computational resources, for example, simulations of computational fluid dynamics (CFD) or MHD to study the cooling of plasma-facing components. As a workaround, in iterative analysis a lower fidelity model may be employed, but this by definition will omit certain effects and limit its accuracy.

%% Data quality
With regards to the second category of data quality, raw or minimally-processed data in fusion are not immediately AI-ready, even when the data release formally adheres to FAIR principles~\cite{wilkinson_fair_2016} (findability, accessibility, interoperability, and reusability). Besides the above issues with limited dataset variety, AI training requires efficient access to slices of the data, and specifically for fusion, domain expertise to inform the sensible ML architectures and choices of predictors and targets.
%As much as a third of enterprise data can be considered redundant, obsolete or trivial (and another 52\% is dark data with unknown value) \footnote{\url{https://www.forbes.com/councils/forbesbusinesscouncil/2023/01/12/how-to-keep-your-data-from-rot-ing-in-the-cloud/}}. How can AI bring value on the large part of trivial data to make scientific discovery? 
%AI can be used for better data storage and there are solutions on AI compression. But how to efficiently interrogate datasets? 
% More work is needed on FAIR datasets, because FAIR data is not always useable. 
%The AI revolution is no different from previous IT revolutions. We need to assess how ready data is for AI. The challenge lies with data management. 

%% Data availability
Finally, in terms of data availability, the majority of experimental and simulation data in fusion is not publicly available. Of course, some of this is due to the sensitivity of data or export control, meaning it is challenging to share data between different jurisdictions. That said, most data in fusion is not sensitive and does not constitute mission critical IP, and the reason for the scarcity of data is mostly due to lack of resourcing for FAIR data initiatives. Foundation models require diverse training data spanning multiple machines, operational regimes, and physical phenomena to achieve the generality that makes them valuable. Community initiatives such as MatDB4Fusion\footnote{\url{https://matdb4fusion.app/}} are beginning to consolidate scattered records into unified databases, but significant effort is still required to make fusion data truly AI-ready with efficient access to appropriate data slices. 

To ensure AI-readiness of fusion datasets such that opportunities are not missed, awareness of FAIR principles amongst the fusion community should be increased.  This may be facilitated through early engagement and collaboration between fusion facilities and data scientists. However, forming collaborations is predicated on the ability to share information;  while navigating issues surrounding security and intellectual property, the ability to cooperate and readily share data through ``science diplomacy'' \cite{Barbarino2021}  will be paramount to making advances in fusion research in general.

%In the context of ``science diplomacy'' \cite{Barbarino2021}, security and collaboration are important aspects to be considered.
%The ability to form collaborations and readily share data will be paramount to making advances in fusion research in general.

\subsection{Validity and Trustworthiness}
Besides data, another challenge for AI is that of validity and trustworthiness. ML algorithms are particularly suitable for interpolation and inference of relationships between fields within large, high-dimensional datasets. These techniques may be employed to organize and classify information in knowledge graphs or to emulate numerical simulations.  However, as such procedures contain no awareness of the underlying mechanisms they cannot be predictive outside their original training domain. Despite this weakness, many AI models will nevertheless make predictions outside their regime of validity with a false degree of confidence, resulting in what is informally known as ``hallucinations''. From a physics perspective, this might correspond to failing to capture some emergent behaviour or phase transitions. 
Therefore, AI approaches should not supersede the development of physical models and understanding of fundamental behaviour by domain experts, but rather augment such knowledge (see also physics-informed approaches in Section 3). In addition, the application of AI methods to fusion would benefit from collaborations involving experts who can provide guidance on the limitations of the training data and how it may be used.

\subsection{Time Scales and Latency}
Although foundation models promise acceleration at inference time, when compared with classical simulation approaches, they require substantial computational resources for training. Critically, for applications in real-time control and power-plant digital twinning (see Section \ref{sec:digitaltwin}), the necessary latency is of the order of milliseconds or shorter—far faster than the seconds required for LLM inference. This motivates the development of specialised "fusion foundation model" architectures that can be fine-tuned and distilled for specific downstream applications while meeting real-time requirements without the latency issues of LLMs, and which can also provide effective real-time data compression, for example, through latent-space representations. On-edge data compression is also needed to reduce the volume of data being streamed for real-time processing. The development of smaller, lightweight models where feasible \cite{zhou2023mini} is receiving increased attention within the EUROfusion community and elsewhere.

\section{Opportunities}

%% Opportunities
Fusion is an inherently diverse area of research, and the problems where AI can be used are just as numerous. In Section \ref{sec:casestudy} concrete examples are given in the context of materials for fusion; AI can help identify and discover new materials, and help guide experiments, narrowing down the large space of possibilities. As an example in a different field, new metal-organic frameworks with promising absorption properties have been discovered thanks to generative-AI techniques \cite{Cipcigan2023}. 

\subsection{Automated Decision-Making}

Control is one of the most difficult challenges for fusion energy, and the control of fusion devices is an area AI is already making an impact. Tokamaks in particular are prone to adverse events such as disruptions, edge-localized modes, and H-L back-transitions; all of which are exceedingly complex phenomena that are difficult to predict from a first-principles basis \cite{vega2022disruption,shousha2025unified,carlstrom2005lh}. AI represents the most effective way to predict and avoid such events, particularly on the often sub-millisecond timescales needed for real-time control \cite{Yang2025,vega2022disruption}. AI techniques such as deep reinforcement learning have also been shown to control such as the magnetic topology of a plasma with great efficacy \cite{degrave2022magnetic}. 

It is uncertain, however,  whether AI for \textit{direct} control of a plasma will be accepted by the operations community, due to issues of trust, transparency and assurance.  The level of risk and impact  that are associated with control failures in operating scenarios is potentially very high: safety, damage to devices, reduction to powerplant availability, financial implications and reputation loss. There are open questions of liability should such failures occur; if a decision is made by AI, who should be made accountable?
Given these uncertainties, in the foreseeable future control systems will likely still require there to be a human ``in-the-loop". 
%Note that the human in the loop is already a reality at the stage of scenario design, so a recommendation would be that this is retained also when real-time control strategies are obtained via AI instead of conventional (and not necessarily more accurate) methods used so far. 
Nevertheless, human intuition is also prone to error; as such, complementing any AI-informed real-time control strategies by 
predictions of plasma behaviour that have confidence intervals obtained through uncertainty-aware toolkits (as discussed in Section \ref{sec:uq}) may more effectively support human decision-making.

As well as control over a tokamak plasma, fusion will require extensive robotics and remote handling hardware \cite{bachmann2022conceptual}. Though these can be controlled manually, this can often make tasks such as the replacement of damaged plasma facing tiles arduous and time consuming. Using AI techniques such as reinforcement learning, or machine-learnt model predictive control could automate many remote handling tasks, significantly decreasing time and resources required for repairs or regular plant assessments \cite{shi2017mechanical}.  Of course, as in the case of plasma control, to ensure safe operations of such tasks, there will remain a need for human supervision. 

\subsection{Accelerating Simulations}

One potentially impactful application of AI in fusion is in the acceleration of detailed physics simulations that are computationally too intensive for practical purposes.

In any multi-scale modelling scenario, of which a notable example for fusion is materials\cite{gilbert_perspectives_2021}, it is not practical to run simulations that range in length-scale from atomistic to macroscopic. If surrogate models are trained to emulate computationally-expensive simulations, these may be employed to connect the emergent properties arising from microscropic physics to determine component-scale behavior. This may aid in assessing life-times and predictive maintenance. 

In a related example, sequential learning methods can be used to efficiently learn a surrogate model of the response surface for indicators of performance derived from expensive computational models, enabling optimization engineering-scale power-plant components\cite{Humphrey2024} and experiments\cite{Humphrey2026}. 

 If pre-trained models are fine-tuned to datasets with smaller size, incomplete information or covering a different domain, issues of transferability and explainability should be considered. In the context of iterative design and parameter explorations, it may be desirable to combine models of both low- and high-fidelity. An AI-enhanced algorithm might recommend within which regions of parameter space it is appropriate to use an existing low-fidelity or surrogate model, and when it is necessary to provide updated data from the high-fidelity model. In this manner, maximal information may be elicited for a given amount of computing time and resource. Such a procedure has already been demonstrated for the design of nuclear reactor cores \cite{sobes_ai-based_2021}.

In these examples the inherent value of a surrogate model is the ability to use it in another context, be that to enable multi-scale or multi-fidelity  modelling, to inform control systems, to accelerate design activities or perform uncertainty quantification (see dedicated section below). It is not suggested to develop surrogate models for their own sake; as the training data need still be generated such endeavours might be better expended in accelerating the original simulations themselves (for example through improved numerical algorithms or modernized to leverage emerging hardware). 

It should also be noted that when developing of surrogate models, it is tantamount that such training is not performed treating the simulation software as a black box.  As noted above, expertise in modelling and simulation must be engaged to ensure limitations of the models are understood.  In the case of optimisation, any figures of merit, parameter constraints and tacit knowledge must be carefully defined and encoded to ensure the optimizer does not converge upon solutions that are practically infeasible. 

%Combining simulation and validation with experiments can be very powerful and AI has a lot of potential in this space.

% AI models have performed well in situations where data is scarce or there is no experimental data and one can use that experience in fusion. We need to be able to learn in one context and then transfer to another. Then data management and security as well as explainability and transferability are very important. 

% \textcolor{blue}{Comment AA: : in non-fusion examples, there are uses of generative AI and graph-based algorithms to infer materials properties(e.g. MOFs for CO2 capture, or ceramics and perovskites).}

 %CC contribution on uncertainty guiding data collection
\subsection{Uncertainty Quantification}
\label{sec:uq}

An important area AI can contribute to is the guiding of scientific advancement through uncertainty quantification. Broadly speaking, much of science is an exercise in uncertainty reduction; the scientific process focuses on identifying the most pressing gaps in our knowledge, and then attempting to fill those gaps with relevant, observable experiments or simulations. AI represents an effective way to formalize this process. Using the variance of data close together in parameter space allows AI models to automatically determine aleatoric uncertainties (driven by noise). ML methods such as GPs can also learn some of the epistemic uncertainties (driven by lack of knowledge) from sparse datasets \cite{williams1998prediction}. This is done by learning how strongly correlated certain input regions are with the training data. Of course, there is also an epistemic uncertainty associated with possibly incorrect assumptions used in simulation tools. These `unknown unknowns' cannot be estimated by such algorithms. Engagement with modelling expertise to advise upon the overall suitability of the simulation tool is necessary to contextualize the uncertainty estimate; in some cases where the models are known to be poor descriptors of data, it might be better to focus upon developing the foundational science itself.

By quantifying uncertainties, AI can also aid in concrete recommendations to reduce uncertainties. This can occur both on the programmatic level, and with low-level decisions. On the programmatic level, quantified uncertainties in data such as material properties can help inform the design of new test facilities. AI methods have also been used to place sensors for observation uncertainty reduction \cite{capellari2016optimal}.  On lower level decisions, AI models can recommend data to sample within a simulation or experimental campaign. Rather than random sampling, AI-recommendations can target new data around reducing the overall system uncertainty, making more efficient use of expensive codes or facilities; particularly those with many free variables \cite{noack2021gaussian}. 

\subsection{Data Collection}

AI models can be used to automatically collect data for other objectives than just uncertainty quantification, through various acquisition functions. AI models can recommend data collection that leads to efficient training and improvement of the models themselves, in a process known as active learning. Active learning has already been shown to make gyrokinetic and exhaust surrogate model training more efficient by more than four times  \cite{zanisi2024efficient,hornsby2024gaussian,Holt2024}. AI models can also recommend sampling data to help with early optimization of component designs, or recommend data which is likely to show a specific type of behavior. For example, finding transition regions in materials testing, or targeting high-success probability regions of often numerically unstable simulations \cite{Agnello2024,chakrabarty2022simulation}

Finally, AI can be used in the management and stewardship of data. AI-based imputation methods enable completion of incomplete or missing data entries. Probabilistic models, anomaly detection models, and language models, may all contribute to interpretation and flagging of possible anomalous data within fusion databases. For example, the U.S Department of Energy Fusion Data Program, led by General Atomics, has collated high quality FAIR data in part using ML techniques from Sapientai \cite{sammuli2024enhancing}.

\subsection{Data Analysis and Interpretation}

AI techniques have demonstrated credible potential to aid the analysis and interpretation of data from fusion experiments. For example, for experimental rigs that are intended to verify and validate simulation tools in fusion. Techniques such as GP regression, or combinations with Bayesian inference, can assess relative sensitivities and uncertainties, allowing for a more robust validation of simulation against experiment. Additionally, these techniques can recommend which measurements will maximally constrain the model or target regions of model disagreement \cite{Jarvinen2022,pavone2023} allowing the optimisation of experiment campaigns for information gain. 

In terms of analyzing data from a fusion plasma, it is incredibly difficult to directly observe plasma parameters with diagnostics. Most plasma diagnostics instead measure corollaries which are then used to infer the non-observable parameters of interest, such as separatrix location. AI can certainly help with this, most obviously by employing Bayesian inference frameworks with machine-learnt forward models \cite{kwak2020bayesian}. However, the analysis of individual complex diagnostics, such as cameras, has become increasingly popular with techniques such as convolutional neural networks \cite{juven2024u,wei2023mhd}.

\subsection{Digital Twins}
\label{sec:digitaltwin}
If the aforementioned techniques were combined such that it was possible to run high-fidelity predictive simulations in real-time (or faster) for a specific physical asset, and continually update the underlying models with experimental data, this could approach the Grieves and Wickers definition of ``digital twin'' \cite{Grieves2017}: \textit{``a set of virtual information constructs that fully describes a potential or actual physical manufactured product from the micro atomic level to the macro geometrical level. At its optimum, any information that could be obtained from inspecting a physical manufactured product can be obtained from its Digital Twin."}
The value to such a twin is broad. As an example, during power plant operations, such a digital twin could minimize downtime required for maintenance, in turn reducing cost of operations and making fusion a more economically viable technology.

% , it can aid in validating simulations through verification, validation, and uncertainty quantification pipelines comparing simulated predictions to experimental observations \cite{roy2011comprehensive}. 

%Quantum computing provides an interesting complement to discoveries that cannot be feasibly achieved with "classical" AI in materials alone. Fusion is one of the few areas where we can see a real difference brought about from using quantum methods \cite{joseph2023quantum}.

%How exactly we define digital twins is different for different problems. The challenge is to develop digital frameworks where different components can communicate with each other. The big challenge is integration and developing digital threads. 

 %[ HB comment  While some people use the terminology "Digital Twin" somewhat loosely, there is a precise definition to be found in this canonical and well-cited paper ]

%additional statements: 

%\begin{itemize}
    %\item Enterprise deployment of fine-tuned models. Applications AI in large industries and manufacturing is a big world of opportunities.
%\end{itemize}

%There are exciting opportunities, but we should be cautious about using the AI hammer, there are problems where AI cannot help. It can provide a first step, but all the hard work thereafter is still needed.

\section{Industry-Academia Collaborations}
The challenges and opportunities discussed above cannot be tackled and exploited without collaboration and long-term engagement between academia and industry. The barriers that may exist to doing so, as well as mitigations, are discussed in detail below.  

\subsection{IP and Publications}
IP is often a barrier in academia-industry collaborations. Although not all IP can be disclosed, the field of fusion tends to be more open than others, both because the community recognises that this is a grand challenge and also because some software and basic science can only be done through large community-wide efforts. In wider industrial R\&D projects, fruitful discussions can be  had early on in the scoping of the projects, and industrial partners sometimes appreciate that some reusability of the assets ends up benefiting them too, trivially because their own R\&D can start from a more advanced stage. 

While academia trains future scientists, industrial placements can be very beneficial not only for developing skills and building bridges between academia and industry. More PhD programs co-supervised by academics and industry partners are needed. However, for such programs to work well the IP should not be too strongly protected so that students can publish.

There is only an apparent tension between publications and impact, since industrial R\&D work can lead to impactful literature, especially in the field of fusion, where there is still a widespread consensus within both academia and industry that discoveries should be shared openly, including the curation and open-sourcing of the associated digital assets such as data and code. Among the participants to the \textit{FusionFest} roundtable, there was unanimous consensus that new and larger cohorts of excellent students should be trained specifically on fusion and its interdisciplinary implications (physics, computation, engineering, project lifetime management), providing occasions of early contact and hands-on work with the outstanding challenges in the field, both within universities and in the industry -- including industrial placements.

\subsection{Skills and Domain Knowledge}

It is worth emphasising that there are many AI techniques and many fusion problems. Not all fusion data will fit into one model, so it is essential to mindfully match specific problems in fusion with specific AI methods.  Students and researchers in traditional STEM programmes may not have sufficient exposure to the latest developments in foundation models for instance, while ML researchers may lack appreciation for the unique challenges of fusion. Attracting researchers from the computer vision and general ML communities—where foundation model expertise is concentrated—requires demonstrating that fusion presents compelling and tractable research problems.

As such,  AI experts and domain experts should be connected to ensure mutual understanding of what can be done and the context in which the solution may be deployed. Furthermore, where many experts are involved in AI for fusion research projects, it can be challenging to maintain this contextual awareness. For this reason fusion-AI collaborations should be maintained  in the long-term to build knowledge and communication so both sides can understand each other's constraints, methodologies, and terminology.

\subsection{Pace and Focus}

It should be acknowledged that academia and industry operate differently. For example, industry typically moves rapidly and must quickly establish the business value of applying AI techniques. Academia may be suitable to test, validate and assess the utility of emerging methods and models; however, the time-frames to do so rigorously might not keep pace with the industrial needs. 

Another example of disparity is the lack of awareness among students and academic researchers about the industry requirements, directions and opportunities. Industry best-practices are not necessarily understood and applied in academia. Where domain experts in niche topics are not AI specialists, there is a need to have access to embedded research software engineers to facilitate transfer-learning of such practices. Here there is an opportunity to improve the quality of implementations. There is value in connecting deep tech companies with engineering applications to create sophisticated tools that design engineers can use. This is something best done embedded in an industrial environment. 
However, with such endeavours the explainability of AI models should not be left to academia alone. There should be more effort from such companies to integrate uncertainty quantification and physics models.

Notwithstanding potential differences in pace, an emerging trend is a shift in focus away from promising academic methods toward more pragmatic approaches that can concretely accelerate the delivery of first-of-a-kind fusion power-plants.

\section{Case Study: the role of AI methods in solving the fusion materials challenge}\label{sec:casestudy}
AI has major potential to contribute to the discovery, development, design and qualification of fusion materials. However, as outlined in section 5, its practical impact is curtailed by one fundamental limitation: the field lacks a sufficiently rich fusion-specific body of materials data.

Machine components for fusion must endure an extreme environment: a 14 MeV neutron flux, 1 to 10 MW m$^{-2}$ heat loads (or greater, in transients), steep thermal gradients, and a complex combination of synergistic mechanical, electromagnetic, chemical and nuclear stresses \cite{CATF}. Existing nuclear materials data stem largely from experience with fission reactors, whose data from sub-2 MeV neutrons and lower fluxes, and with single-physics effects, provide only a limited surrogate for fusion conditions. As a result, the degradation mechanisms most relevant to fusion cannot be reliably inferred from fission experience. A deuterium–tritium‑fueled fusion power plant can see material damage rates approaching tens of displacements‑per‑atom (dpa) per year, together with substantially higher helium‑per‑dpa transmutation levels—phenomena rarely encountered in current nuclear systems—so swelling, embrittlement and conductivity‑loss pathways may diverge from those predicted by simple extrapolation of fission data \cite{zinkle2009structural}. These coupled effects severely limit component lifetime, meaning that designers must rigorously quantify degradation rates when developing breeding blankets and planning maintenance intervals. Consequently, predictive models for fusion face the classic “Garbage In, Garbage Out” (GIGO) constraint: without fusion‑relevant, high‑fidelity inputs, even sophisticated algorithms will struggle to produce reliable forecasts.

Adding to this challenge, the successful development of fusion energy does not depend on a single “miracle material” but on an array of materials, each required to withstand distinct combinations of extreme loads. First‑wall armour, structural supports, breeding blankets, divertors, superconducting magnets, and even laser‑facing components each have specific performance criteria that must be met. Tens of material systems therefore require development and qualification, and their readiness diverges strongly depending on whether the goal is experimental (e.g., ITER‑class) operation or commercial conditions, where lifetimes, fluences, and functional demands—such as full tritium breeding—are significantly more stringent.

Today, many candidate materials for fusion remain at relatively low technology readiness levels, generally around TRL 3-4 \cite{alba2022materials}. ITER has opted for comparatively mature materials choices:  a tungsten divertor, a tungsten first wall (in place of the beryllium originally planned), and reduced‑activation ferritic‑martensitic (RAFM) steel in the structural components. Although these materials have been developed in parallel with ITER’s decades‑long design and construction, they will still be pushed to their operational limits once ITER begins sustained running. Their performance is therefore unlikely to meet the demands of a next‑step commercial prototype, which will require longer operating lifetimes and more stringent capabilities, including full tritium breeding. There, higher-performance candidate materials such as silicon carbide composites, vanadium alloys, and oxide-dispersion-strengthened (ODS) steels show great promise. However, they remain largely at laboratory scale or are produced only in small batches for industrial testing, and are therefore far from being developed to the level required for deployment in a fusion power plant \cite{alba2022materials}. Moreover, the volume of fusion‑relevant irradiation data for these advanced systems is even more limited.

So, fusion developers must decide whether to rely on “good-enough” industrial alloys---mostly fission or aerospace grades with abundant data, but likely to survive for only a limited lifetime in a fusion environment---or to invest in entirely new radiation-tolerant materials, such as high-entropy alloys, which promise superior performance, with no guarantee of eventual success and at the cost of lengthy R\&D and qualification. In practice, both paths must advance in parallel. Here, AI offers a means of introducing greater structure and rigour into the assessment of candidate materials. Through quantitative analysis of legacy datasets and published results, data‑driven methods can determine which existing alloys have demonstrable evidence supporting their prospective performance and, conversely, where the underlying data are incomplete or insufficient to justify further design decisions. That insight helps direct the allocation of expensive and time-intensive experiments, including irradiation tests and prototype trials. At the same time, compututational algorithms can rapidly scan the vast combinations of elements and processing routes to suggest new, untested alloys worth exploring in the laboratory---allowing researchers spend time only on the most promising candidates, while still relying on real experiments for definitive validation.

One such example is that Microsoft researchers recently used a coupled model-simulator workflow to demonstrate dramatic acceleration in the identification of ultra-strong, low-activation alloy candidates--reducing what would typically require years of iterative exploration to the order of weeks \cite{ITER}. Another example comes from Oak Ridge National Laboratory (USA) where AI methods coupled with exascale computing were used to identify high-temperature, radiation-resistant alloys, thereby avoiding much of conventional trial-and-error that characterises traditional materials development programmes \cite{OakRidge}. Together, these demonstrations show how algorithmic search can accelerate innovation across a design space far beyond the reach of human intuition or traditional experimentation. In doing so, they help reduce the risk associated with current materials choices while clearing a viable path to the superior higher-performance materials that commercial fusion will ultimately demand.

Whether repurposing known alloys or inventing new ones, the target performance envelope must first be defined, yet these targets remain open-ended. Beyond ITER-class devices, most pilot-plant designs are still at an embryonic stage \cite{cohen2024long}; engineers cannot yet confidently specify concrete requirements, such as “we need an alloy that survives $X$ dpa, at $Y$ °C, for $Z$ years” because $X$, $Y$, and $Z$ exist only as broad ranges or aspirational goals. Acceptable limits on things like radiation damage, temperature and stress are greatly dependent on the eventual fusion machine architecture—--which itself is very likely to evolve in a coupled manner as materials constraints (and other system constraints) emerge. Moreover, looking at fusion more broadly, these limits and the requirements for materials will also differ markedly between magnetic, magneto-targetised, and inertial confinement system approaches.

This creates a classic chicken-and-egg problem. Cohen-Tanugi et al. advocate an iterative co-evolution of materials R\&D and machine design, in which requirements and candidate solutions are refined in tandem \cite{cohen2024long}. Such a systems approach concedes that ideal materials will not be available at the outset but instead that plant design and materials development must continuously inform one another, with engineers relaxing design constraints where feasible, while materials scientists push toward more ambitious configuration envelopes.

AI can ease the dearth of data that underpins this chicken-egg problem in two complementary ways. First, it can extract substantially more value from the limited datasets that are already available. ML methods can systematically analyse the extensive body of fission‑irradiation experiments, accelerator‑driven surrogate studies, and laboratory materials tests—together with the limited fusion‑spectrum data that do exist—to identify statistically robust trends that are difficult to extract through manual review alone. Community projects such as MatDB4Fusion are now consolidating these disparate records into a unified database, enabling models to predict fusion-like behaviour at least within observed bounds \cite{CATF}. Early work by Kemp et al. (2006, \cite{kemp2006neural}) showed that neural networks could capture irradiation-hardening trends in low-activation steels that physics models alone could not. Modern approaches extend this by incorporating physics-informed constraints and providing uncertainty-quantified predictions for doses or temperatures beyond those directly tested. This gives experimentalists a more structured, data-driven “map”, of sorts, for anticipating material behaviour rather than relying solely on extrapolation or intuition alone.

AI‑based design‑of‑experiments methods can prioritise alloy–dose–temperature combinations by estimating which candidate test would yield the highest information gain. This approach directly addresses the perennial question: “If only one more irradiation could be run, which one would reduce uncertainty the most?” Such optimisation is not optional but structurally necessary for facilities like IFMIF‑DONES—--a high-energy 14 MeV neutron source being built in Granada, Spain, intended to generate fusion‑relevant benchmark data—--because they will operate with high cost and limited throughput for years, meaning judicious planning and materials selection is vital \cite{alba2022materials}. Real-time analytics can maximise each shot’s value—refining test matrices promptly and flagging likely winners and losers. In short, AI acts as a compass for fusion materials R\&D: identifying statistically meaningful patterns in existing datasets, prioritising experiments according to expected information gain, and accelerating progress in domains where fusion-oriented data is sparse at least until fuller datasets arrive.

In summary, although progress toward fusion‑ready materials ultimately relies on rigorous irradiation campaigns and high‑resolution microstructural characterisation, these experimental programmes can be materially enhanced by AI‑enabled computation and modelling. AI cannot create data that do not exist, but it can extract far more value from what is available: filtering noise, exposing correlations, quantifying uncertainty, and recommending the next experiments that would be most informative. When ML tools are combined with HPC and laboratory automation, the field can shift from traditional “test-analyse-repeat” workflow to an iterative, data-driven loop. This could potentially shorten the path from concept to demonstration, and then qualification to just a few years—which could prove vital for pilot plants targeting first power within a decade \cite{CATF}.

The obstacles, however, remain formidable: fusion neutrons will still drive material degradation including swelling, embrittlement and transmutation. Datasets capturing these phenomena under fusion‑spectrum conditions remain sparse. This “fusion materials gap” of sparse data, evolving performance targets (based on fusion designs) and altogether low-TRL will not be eliminated quickly. Yet AI offers a systematic way to make progress despite these constraints—--by de-risking designs, tightening material property envelopes, and highlighting regions of composition or processing that merit focused investigation. By combining pattern‑recognition tools with domain expertise, researchers can direct scarce irradiation time toward the experiments that matter most. Real-world materials science and engineering remains central; AI merely serves as an analytical lens that sharpens insight and accelerates decision-making, helping a quest once measured in decades to be instead tackled in the order of years\cite{ITER}.

\section{Summary discussion}

This Perspective emerged from the discussions at The Economist \textit{FusionFest} roundtable held in April 2025. The event was attended by 29 representatives from academia, UKAEA, STFC, industry and start-ups who all contributed to the discussions. A subset of the participants developed the main points further in light of the fast-evolving technological and socio-political context. The original discussion was extensive and in this Perspective the authors focused on the most important themes summarized below:
\begin{itemize}
\item	\textbf{There is no one AI model to solve fusion in general}. There are many types of AI tools and many types of fusion problems. The challenge is how to match problems with the right AI tools which requires domain expertise in both fusion and AI methods that can only be achieved through better communication and collaboration.
\item	\textbf{Data is needed to train AI but data is scarce in fusion}. AI methods can help guide experimental data taking to the regions of most interest and simulations to the parts where highest accuracy is needed. These are more efficient and effective ways to produce real and synthetic data.
\item	\textbf{Collaborations between fusion domain experts and AI experts are a necessity}. There are already some good examples of how academia and industry can work together, but there is more to be done.
%\item	There are opportunities to advance fusion outside AI-enabled breakthrough in R\&D. In particular in the regulation space AI can be very effective if done right.
\item	The participants seemed to agree that \textbf{it is important to use AI mindfully}, that is, in an explainable and trustworthy way and recognising that \textbf{not everything can be solved by AI, but rather AI can help deploy resources more efficiently}.
\end{itemize}

\begin{table}[htbp]
\centering
\caption{Synthesis of challenges and opportunities in AI for fusion research.}
\label{tab:ai_fusion_summary}
\renewcommand{\arraystretch}{1.2}
\begin{tabularx}{\textwidth}{p{3cm} X X X}
\hline
\textbf{Application} &
\textbf{Opportunities} &
\textbf{Challenges} \\
\hline

Plasma physics \& control &
Disruption and instability prediction; sub-millisecond real-time plasma control; control of magnetic topology; improved operational regimes &
Trust and acceptance by operators; limited interpretability; risk of out-of-distribution predictions and hallucinations; latency constraints for foundation models \\

Materials \& components &
Accelerated alloy discovery; prioritisation of irradiation experiments; uncertainty-quantified prediction of material properties; faster materials qualification and TRL progression &
Severe scarcity of fusion-relevant data; reliance on fission surrogate datasets; evolving performance targets; uncertainties in ground truth; relatively low TRLs \\

Engineering \& design  &
Rapid parameter sweeps; component and system optimisation; coupling of multi-scale models; faster design iteration \ &
Generalisation uncertainty outside training domain; high training cost; explainability challenges; fidelity of underlying simulations \\

Diagnostics \& data &
Improved inference of hidden plasma parameters; faster analysis of heterogeneous data with diagnostic drift; uncertainty-aware data interpretation; digital twins  &
Limited ground truth; diagnostic noise; extrapolation risks; uncertainty sources that cannot be learned by ML\\

Operations \& maintenance &
Automated remote handling and robotics; predictive maintenance; reduced downtime; safer operation in high-radiation environments; digital twins  &
Deployment trust and safety assurance; validation of AI-driven actions \\

\hline
\end{tabularx}
\end{table}
It came across from the discussions that there were multiple problems where AI can be used to advance fusion: in accelerating fusion materials research, in advancing fusion plant design, in scaling up simulation capacity to give just three examples. There are opportunities for academia and industry to collaborate on fusion research and extend their respective fields of expertise. For example, academics can broaden their understanding of the wider fusion landscape and industry can develop and refine uncertainty quantification methods.

There was consensus that the full benefit of AI will only be realised with informed use. This requires skills development, long-term collaboration between fusion and AI experts, good data management and curation practices, robust scrutiny, and understanding of uncertainties.  The best use of AI requires us to apply the right focus --- identifying which problems AI can help with and which AI tools are most appropriate to address those problems. Adopting and embedding FAIR data principles will also help with robustness and trust that results are reliable.

It is worth highlighting that, as in many other experimental fields, there is a large amount of grey literature and tacit knowledge (not always documented, see \cite{barabaschi2024importance}) that enables research. This is why the fusion expertise of scientists, engineers, and technicians is essential in the development and use of AI models.

%\textcolor{blue}{[Add two sentences on recommendation for prioritisation. Refer to the published fusion strategies (e.g. UKAEA's) and highlight the need for twining capabilitites to accelrate the predictive design of facilities, so the priorities of AI-for-fusion initiatives would need to show how they can enable faster experiment and discovery cycles, resolve computational bottlenecks, and aid the development and integration of digital twin components.]}

With such a breadth of potential challenges and opportunities, given a finite set of resources (e.g., skilled people and funding)
it is reasonable to consider how best to prioritise activities. 
We prefer to avoid a set of specific recommendations that may be partial to the composition of the panel and the authors, but some general themes can be outlined as motivated above and in the recent publication of fusion strategy documents published by the UK government and UKAEA~\cite{UKFusionStrategy, UKAEAStrategy}, of which we distil some salient points here. 

AI is a prominent theme in both documents, most ostensibly through the investment in the new, fusion-dedicated \textit{SUNRISE} AI-supercomputer. Overcoming data challenges is highlighted: a notable near-term strategic objective to support design of the Spherical Tokamak for Energy Production (STEP) includes the need to establish suitable data infrastructure; the need to ensure fusion data are consistent, accessible, and electronically readable is emphasized; the generation of datasets for fuel cycle and fusion technology to support digital exploitation is specified as an intended technical outcome. These activities should precipitate the application of burgeoning AI methodologies, such as those discussed in Section \ref{sec:methods}. While numerous categories of opportunities for AI are identified (including surrogate modelling, operations and maintenance), plasma and materials science are particular domains for which flagship technical objectives relating for AI (due by 2030) are associated. The need to select suitable methods, and to establish collaborations between academia and industry is explicitly reflected, echoing the themes raised above.

Within this context, any prioritisation of AI activities should align with the need for actionable digital twins, which can run in (close to) real time, to accelerate the design and discovery cycle. Cross-collaboration is essential to ensure that higher TRL is reached, by developing solutions that can eventually be applied within the multi-scale and multi-fidelity specifications of power plant operations.

% \textcolor{blue}{Final closing sentence?}

\section{Acknowledgments and Contributions}
 The roundtable was moderated by Tara Shears and Melanie Windridge; the summary was compiled and edited by Iulia Georgescu; Richard Pearson contributed the case study in section 8. Alex Higginbottom contributed section 3. Adriano Agnello, Helen Brooks and Cyd Cowley updated and expanded sections 4-7. All the authors participated in the roundtable and then contributed to the development of the original discussion points, edited and reviewed the current version of the article. 
 The authors are grateful to Jonathan Graves for inviting this article and providing valuable guidance and comments.

\section*{References}
\bibliography{refs}

\end{document}